# Prediction of Chronic Kidney Disease Using Deep Neural Network

Iliyas Ibrahim Iliyas[1], Isah Rambo Saidu[2], Ali Baba Dauda[3]
and Suleiman Tasiu[4]

[1,2]Nigerian Defence Academy, Nigeria
[3]University of Maiduguri, Borno State, Nigeria
[4]Federal University Dutsinma, Katsina State, Nigeria,

**Abstract -** Deep neural Network (DNN) is becoming a focal point in Machine Learning research. Its application is penetrating into different fields and solving intricate and complex problems. DNN is now been applied in health image processing to detect various ailment such as cancer and diabetes. Another disease that is causing threat to our health is the kidney disease. This disease is becoming prevalent due to substances and elements we intake. Death is imminent and inevitable within few days without at least one functioning kidney. Ignoring the kidney malfunction can cause chronic kidney disease leading to death. Frequently, Chronic Kidney Disease (CKD) and its symptoms are mild and gradual, often go unnoticed for years only to be realized lately. Bade, a Local Government of Yobe state in Nigeria has been a center of attention by medical practitioners due to the prevalence of CKD. Unfortunately, a technical approach in culminating the disease is yet to be attained. We obtained a record of 400 patients with 10 attributes as our dataset from Bade General Hospital. We used DNN model to predict the absence or presence of CKD in the patients. The model produced an accuracy of 98%. Furthermore, we identified and highlighted the Features importance to provide the ranking of the features used in the prediction of the CKD. The outcome revealed that two attributes; Creatinine and Bicarbonate have the highest influence on the CKD prediction.

## 1     Introduction

Machine Learning is the scientific field dealing with the ways in which machines learn from experience. For many scientists, the term "machine learning" is identical to the term "artificial intelligence", given that the possibility of learning is the main characteristic of an entity called intelligent in the broadest sense of the word (Joshi and Chawan, 2018). The purpose of machine learning is the construction of computer systems that can adapt and learn from their experience. With the use of machine learning to extract useful information can be lend in solving various problems, kidney disease included. Purusothaman and Krishnakumari, (2015) indicated that kidney failure falls one among several classes of disease such as heart disease, blindness etc. which results due to chronic diabetes. Dialysis and transplant are the only method to keep the kidneys function artificially and it is also painful and expensive process. According Luyckx, and Stanifer, (2018) kidney disease increased globally from 19 million in 1990 to 33 million in 2013 in 2016 and in 2010 2.62 million people received dialysis worldwide and the need for dialysis is projected to double by 2030.

In Nigeria, the situation is such that Chronic Kidney Disease (CKD) represent about 8-10% of hospital admissions (Ulasi and Ijoma,2010). Therefore, diagnosis technique is needed so that control or precautions can be taken before becomes late. To acquire important information from medical databases, techniques from various data mining was found very much useful (Chahal and Gulia 2019). By combining machine learning and statistical analysis, very useful information can be drawn from medical databases.

Machine learning methods coordinates various statistical analyses and databases helps to extract hidden patterns and relationships from huge and multiple variable data. The accuracy of a given classifier is ensured through testing the model or technique. Moreover, attributes like Specificity, Sensitivity, and Accuracy are common for disease detection (Padmanaban and Parthiban 2016).

In this paper we used a Deep Neural Network model to predict CKD and evaluate the performance of the model by computing the Specificity, Sensitivity, Recall, ROC Score, Kohens Kappa, F1 Scoreand Precision.

The rest of the paper is organized as follows: Section 2 provides the related work on ML and DNN. Section 3 presents how the research is carried out, while Section 4 provides the results of the research and the discussion. In Section 5, we provided the conclusion of the paper, recommendations and direction for further work.

## 2   Related Work

A remarkable number of researches have been conducted in the area of machine learning for building models that assist in predicting different types of different types of diseases and health related problems, using different machine learning algorithms. This section presents a review of some of conducted research in the area of machine learning in Chronic Kidney disease prediction.

Ge et al. (2019) Predicted parkisons disease severity using Deep Neural Network with UCI's parkison's telemonitoring voice dataset of patients. The studies comprised a biomedical voice measurement of 42 patients with Parkisons Disease (PD). Severity prediction on the basis of total Unified Parkisons Disease Rating Scale (UPDRS) accuracy score of 94.4422% and 62.7335% for train and test dataset respectively and severity PD severity on the basis of total UPDRS accuracy score of 83.367% and 81.6657% for train and test dataset respectively.

Ayon and Islam (2019) Proposed a strategy for the diagnosis of diabetes using DNN on PIM Indian Diabetes (PID) dataset from UCI machine learning repository with an accuracy of 98.35%, F1 Score:98% and MCC:97% for five-fold cross validation. Additionally, accuracy of 97.11%, Sensitivity :96.25% and Specificity:98.80% obtained for ten-fold cross validation and indicated that five-fold cross validation showed better performance.

Shafiet al. (2020) Proposed a machine learning based solution to avoid cleft in the mother's womb with Deep Leaning method and other four methods, on 1000 pregnant female samples from 3 different hospitals in Lahore, Punjab. The authors performed data cleaning, scaling and feature selection method and compared the accuracy for all the algorithms with Random Forest(RF) algorithm:85.77%, Decision Tree(DT):88.14%, K-Nearest Neighbor (KNN):89.72%, Support vector Machine(SVM):90.69% and Multilayer perception(MLP) which is a Deep Neural Network:92.6%. and indicated that MLP yield a better accuracy.

Sharma and Parmar (2020) Proposed a model for heart disease prediction with DNN model on heart disease UCI dataset with six (6) different classifiers KNN, SVM, NB, RF and DNN using talos optimization. Their work indicated an accuracy for KNN:90.16%, Logistic Regression:82.5%, SVM:81.97%, NB:85.25% and DNN with Talos optimization:90.78%.

Ahmed and Alshebly (2019) applied different machine learning algorithm which are Artificial Neural Network and Logistic Regression (LR) to a problem in the domain on medical diagnosis and analyzed their efficiency of the prediction on 153 case and 11 attribute of CKD patients, the observed performance of the ANNs classifier is better than LR mode with the accuracy of 84.44%, sensitivity of 84.21, specificity of 84.61% and Area Under the Curve (AUC) of 84.41% and found that the most important factors that have a clear impact on chronic kidney disease patients are creatinine and urea.

Chimwayi et al. (2017) applied neuro-fuzzy algorithm to determine the risk of CKD patients, using a dataset which had 25 attributes (14 nominal and 11 numeric) describing early stages of CKD in Indians, with accuracy of 100%, sensitivity of 100% and specificity of 97%, they suggested that the work should be added to the domain of healthcare and can be used for providing suggestions in the domain by making it easy for healthcare professionals in diagnosis and treatment of patients as well as for identifying relationships within diseases suffer by patients.

Kriplani et al. (2019) studied 224 records of chronic kidney disease available on the UC Irvine (UCI) machine learning repository named chronic kidney diseases dating back to 2015 and proposed an algorithm. Their proposed method is based on deep neural network which predicts the presence or absence of chronic kidney disease with an accuracy of 97%. Compared to other available algorithms, the model built shows better results which is implemented using the cross-validation technique to keep the model safe from overfitting.

Norouzi, et al. (2016) presented an Adaptive Neuro Fuzzy Inference System (ANFIS) for the prediction of renal failure time frame of KD on a 10-year real clinical data of diagnosed patients, the dataset had 10 attributes, ANFIS model was used for estimating GFR at subsequent 6, 12, or 18 months. Their model achieved an accuracy of 95%, urine protein was not used among their parameters.

Scholar (2018) employed Decision Tree (DT) and Naïve Bayes(NB) as machine learning algorithm to predict Chronic Kidney Disease (CKD) using UCI machine repository dataset with 25 attributes and achieved an accuracy of 99.25% and 98.75% for DT and NB respectively, showing DT as better algorithm in terms of predicting the presence and absence of CKD.

Arafat et al. (2018) studied an automated detection of CKD with clinical data using Random Forest (RF), Naïve Bayes(ND) and Decision Tree based on a comparative study on UCI dataset, they computed the weight of each attribute used in the dataset. Their result shows that RF has higher accuracy of 97.5% and 96% for Naïve Bayes and Decision Tree in the prediction of CKD.

Based on the literature reviewed, several studies worked on the presence and absence of chronic kidney disease in a comparison process and prediction of CKD using various AI Techniques, but less work has been done with DNN model in the prediction of Chronic Kidney Disease to achieve a better result.

## 3    Methodology

This methodology depicts the design of the method to be exploited to carry out the experiment. It incorporates data collection, data preprocessing, and target variable selection, prediction DNN model and performance evaluation as shown in Figure 1.

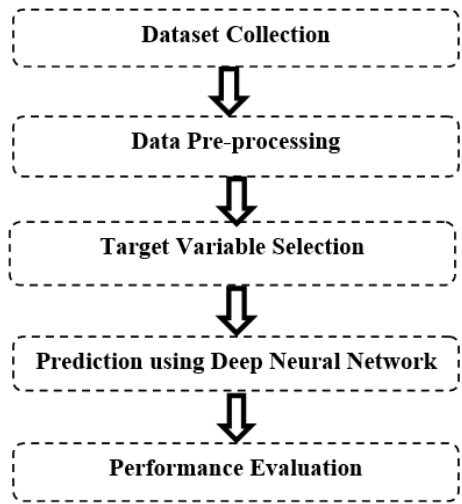

Figure 1: Flow of Research Methodology

**Data Collection**

Patient's kidney disease record is selected as the source of data for this work.This dataset is collected from General Hospital inGashua Local Government Area of Yobe State. It contains 400 patients record with 11 attributes/parameters: Age, Gender, Sodium, Potassium, Chloride, Bicarbonate, Urea, Creatinine, Urea Acid, Albumin and Classification including a target variable classified into a binary classification of CKD and non CKD disease as shown in Table 1.

Table1: Dataset Attributes

| Attribute | Attribute Description |
|---|---|
| Sex | Gender |
| Age | Age |
| Sod | Sodium |
| Pot | Potassium |
| Chl | Chloride |
| Bica | Bicarbonate |
| Urea | Urea |
| Cre | Creatinine |
| UA | Urea Acid |
| Alb | Albumin |
| Class | {Kidney Disease, NoKidney Disease} |

**Data Pre-processing**

Data Pre-processing represent the most important task in data mining techniques, it involves cleaning, extraction and transformation of data into a suitable format for machine execution,

Raw data contains missing information, bad formats, invalid information and it leads to disaster in prediction with machine learning. The dataset used had some missing cells which was replaced using simple imputation with the mean value of the attribute and the attribute Sex was converted to numeric values as '1'and '0' for Male and Female respectively to make it possible for the machine to process since the machine will not understand string values.

**Target variable**
CKD dataset has got a lot of useful variables which are key and necessary for the identification of the disease in patients.we used our variables based on the test and method used by the Hospital in determining the occurrence of kidney disease which is Ten (10) blood test and that are the once we used as our input variables. Names too was among the test, but we decided to remove it since it has no impact on our test and the privacy of the patients must be kept.

**Prediction using DNN**
DNN is a subset of Artificial Neural network which simulate the structure and functionalities of biological neural network consisting of an input, weights and activation function, the structure of DNN has an input, hidden layer and an output.In DNN, a is referred to the output, where Wi and Xi are the weight and input respectively. This is represented as:

We used model has ten (10) inputs, two (2) hidden layers and one (1) output in the adopted model. Input layer is set of neurons which take the actual input data for processing. The number of neurons can be decided in accordance with the input data. To train the model, three layers were used. For input and the hidden layer, "Relu" is used as activation function. Relu outputs in 0 or 1. The output layer has only one output result either KD or NKD with Sigmoid as activation function. Stochastic Gradient was used as the optimizer of the model. The experiments are constructed with Python Programming Language. Figure 2 shows the structure of the DNN model adopted for this study.

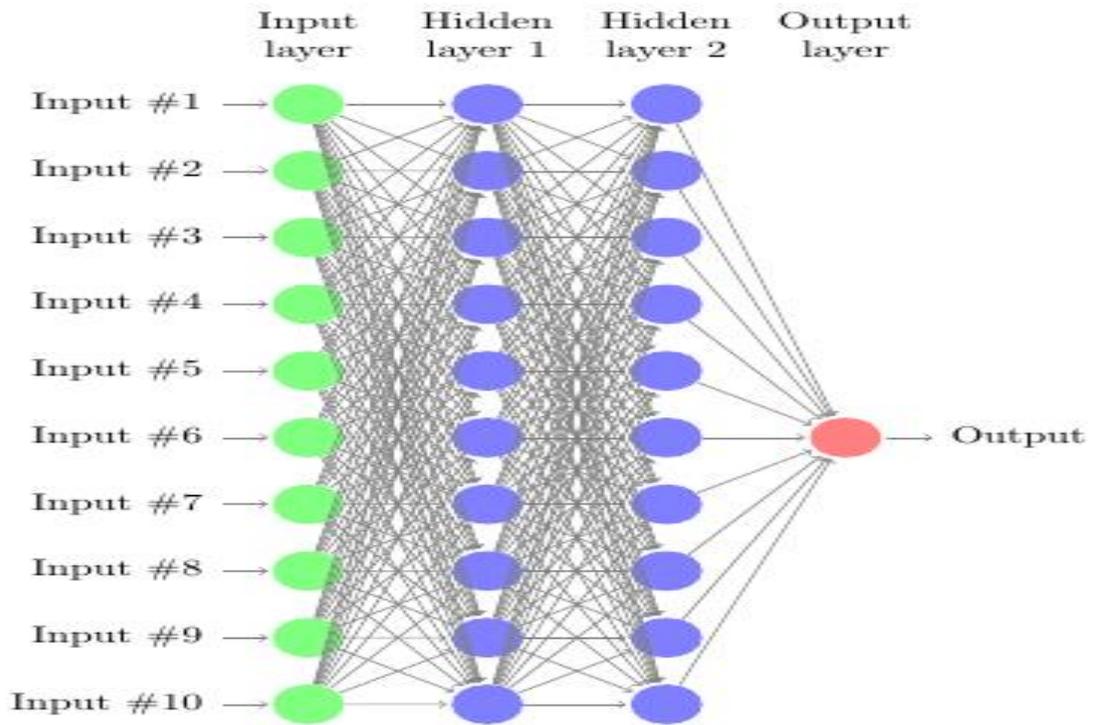

Figure2: Deep Neural Network

**Evaluation of The Model**
In this work, the performance is measured by Accuracy, specificity, sensitivity, kappa statistic, precision, F1 score, ROC Score and recall described as follows.

- Confusion Matrix: confusion matrix indicates statistical suitability of the model and its compatibility with the dataset; it can also be defined as a table layout that is specifically used for visualization of algorithm performanceTable 2 shows the summary of confusion matrix.

Table2: Confusion Matrix

| Classification | | Observation | |
|---|---|---|---|
| | | Negative | Positive |
| Observations | Negative | True Negative(TN) | False Positive(FP) |
| | Positive | False Negative(FN) | True Positive(TP) |

- Accuracy- it is used to identify the number of correctly predicted data points out of all data points. It is defined as the number of all correct predictions made divided by the total number of predictions made, it is expressed as;

- Sensitivity- (Recall or True Positive Rate):it is defined as the proportion of actual positive cases that got predicted as positive. It is a ratio of true positive to the sum of true positive and false negative. In medical diagnosis, test sensitivity (Recall) is the ability of a test to correctly identify those with the disease, it is expressed as;

- Specificity- (True Negative Rate): it is defined as the proportion of actual negative that got predicted as the negative. it is calculated as the number of correct negative predictions divided by the total number of negatives. It is also called true negative rate, it is expressed as;

- Cohen Kappa: This is a classifier performance measure between two sets of classified data. Kappa result values are between 0 to 1. The results become meaningful with increasing values of kappa. it measures how closely instances classified by the machine learning classifier matched the data labelled as the truth, it is expressed as;

- Precision: It is defined as the fraction of relevant instances among the retrieved instances. This is given as the correlation number between the correctly classified modules to entire classified fault-prone modules, it is expressed as;

- Recall/ Sensitivity:Recall is a metric that quantifies the number of correct positive predictions made out of all positive predictions that could have been made, it is expressed as;

- F1 Score: This is the harmonic mean between precision and recall. Range for f1-score is from 0 to 1. It describes the preciseness (how many records can be correctly classified by the model) and robustness (it avoids missing any significant number of record) of a model. The expression of F1-score is as follows;

# 4 Results and Discussion

4.1 Results

This Section provides the results of the DNN algorithm produced based on the dataset.

Table 3: Confusion Matrix for Deep Neural Network

| Classification | | Prediction | |
|---|---|---|---|
| | | Absence | Presence |
| Observation | Absence | 29 | 0 |
| | Presence | 1 | 30 |

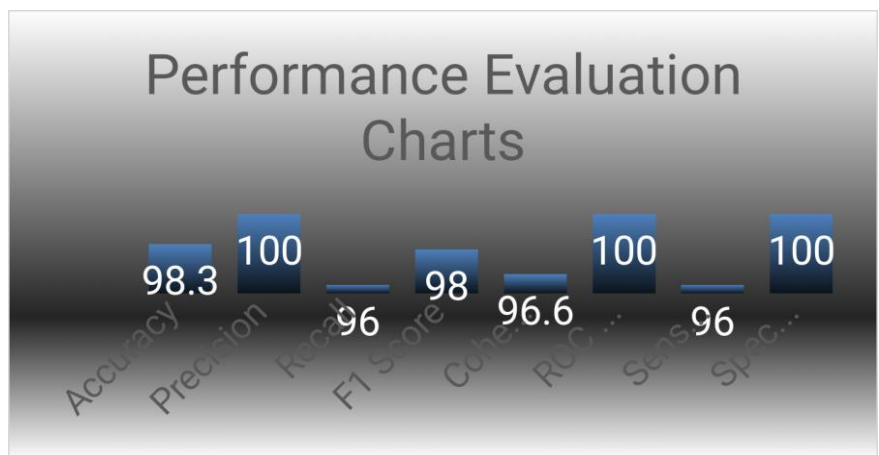

Figure 3: Performance evaluation for Deep Neural Networks Model

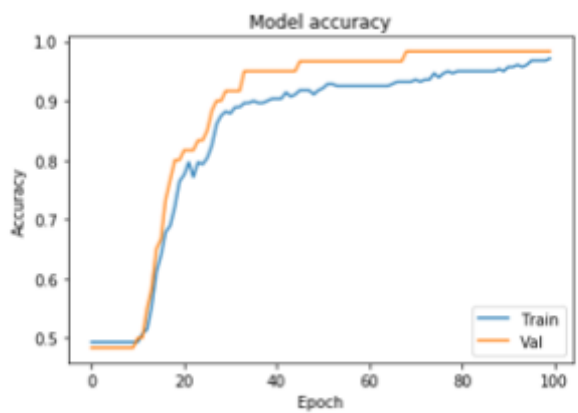

Figure 4: Graph of training against Validation Accuracy

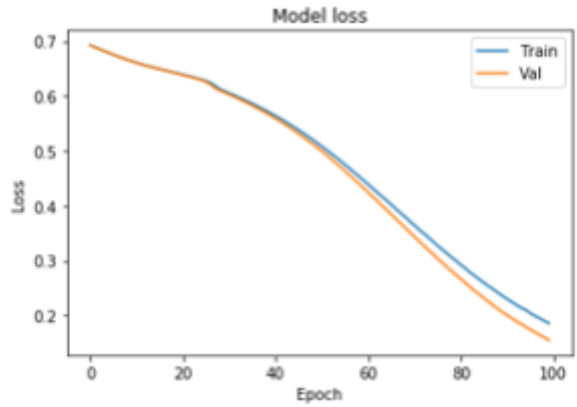

Figure 5: Graph of Training against Validation Loss

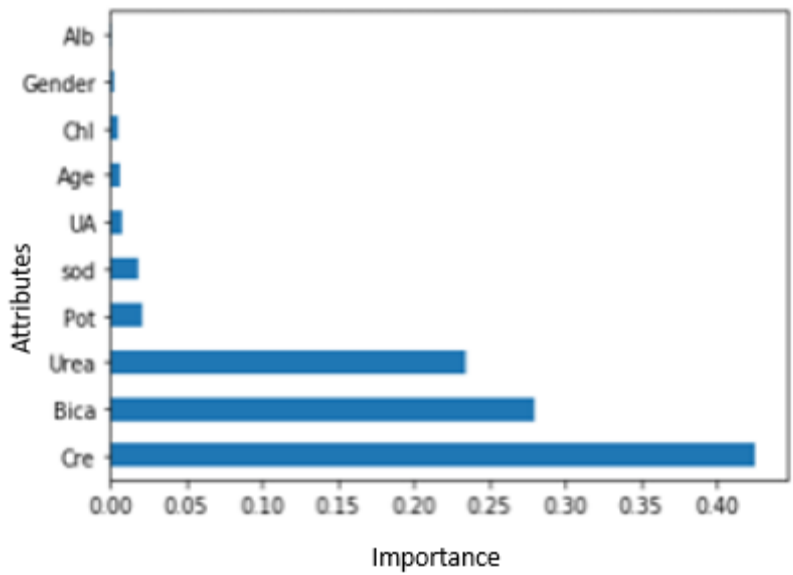

Figure 6: Feature Importance

**4.2 Discussion**

We divided our dataset into 70% (280) and 30% (60) for testing and training respectively, Table 3 shows the confusion matrix provided a visualization summary of DNN model on the dataset, the model has succeeded in classifying 34 samples correctly which are non CKD and 25 samples correctly which are CKD but 1 sample which is CKD was classified wrongly. Totaling to 59 samples classified correctly indicating that the DNN model perform well on our dataset since only one sample out of the sample selected for testing was not classified.

Figure 3 represents performance evaluation of the model and it indicate that the model had succeeded in predicting kidney patients record into CKD and non-CKD with 98% overall prediction accuracy when the model was tested with an unseen data. Based on this accuracy of 98% the model indicated that it is hyper intelligent to seamlessly make prediction into CKD and non-CKD. Other metrics used to evaluate the performance of the DNN model are Precision :100%, Recall:96%, F1 Score:98%, Cohens Kappa:96.6%, ROC Score:100%, Sensitivity:96% and Specificity:100%.

Figure5 shows both validations set and training set accuracy loss of our model as it iterated 100 times and based on the result in the figure. This shows no overfitting since there is improvement because as it iterates the loss accuracy decreases in both the training and testing sets with almost the same level of declining.

Figure 4 shows the level of our model accuracy as it iterated in both training set and validation set and the more it iterates the more the accuracy increases in both sets and with almost the same level of advancement with the validation set even surpassing the training set indicating effectiveness of our model and no overfitting.

Figure 6shows the ranking of features in the dataset that has more influence in CKD prediction from the 10 attributes, Creatinine and Bicarbonate has the highest ranking followed by Urea, Potassium, Sodium, Urea Acid, Age, Chloride, Gender and Albumin, Creatine and Bicarbonate has much influence during the prediction with the DNN model.

## 5    Summary, Conclusion and Further Research

This Section provides the summary of the paper, the conclusion by identifying performance of the DNN and the attributes that were found too pronounced. The Section also enumerates some research that can be done in future to provide effective performance.

### 5.1 Summary

This work collected and trained DNN model to predict the absence and presence of Kidney Disease with an accuracy of 98% and other seven (7) metrics were used to evaluate the performance of the adopted modeland provided a summary of attributes in the dataset used that influence Chronic Kidney Disease (CKD) prediction.

## 5.2 Conclusion

The main goal of this research is to use DNN model for the prediction of kidney disease to high degree of accuracy. We succeeded in classifying kidney disease dataset into CKD and non-CKD with 98% overall accuracy when the model was tested with a set of data that were not used during the training process. The adopted DNN model proved to be efficient and suitable for the prediction of kidney disease. The study also highlighted the importance of the features used in the prediction of kidney disease. This revealed that from the 10 attributes, Creatinine and Bicarbonate are the attributes with highest influence on CKD as opposed to the findings by Ahmed and Alshebly (2019) that found that Creatinine and Urea has the highest influence on CKD prediction.

## 5.3    Further Research

Incorporating more data to have large dataset will provide more accuracy and efficiency; hence more dataset is needed to accommodate enough samples. Having enough sample will make the prediction wider to capture and identify regions and areas with CKD vulnerability. This paper used DNN model only, having other models to compare techniques may provide a better performance.